\documentclass[]{eusset}
\usepackage{listings} 
\usepackage{hyperref}
\usepackage{booktabs}
\title{What Did They Mean? How LLMs Resolve Ambiguous Social Situations across Perspectives and Roles}

\author{Qiming Yuan}
\email{qyuan2@scu.edu}
\affiliation{
  \institution{Santa Clara University}
  \country{USA}
}
\author{Linyi Han}
\email{linyihab25@gmail.com}
\affiliation{
  \institution{Santa Clara University}
  \country{USA}
}

\author{Nam Ling}
\email{nling@scu.edu}
\orcid{0000-0002-5741-7937}
\affiliation{
  \institution{Santa Clara University}
  \country{USA}
}
\author{Cihan Ruan}
\email{luciacihanruan@gmail.com}
\orcid{0009-0006-3094-0505}
\affiliation{
  \institution{Santa Clara University}
  \country{USA}
}

\copyright{CC-BY-4.0}

\begin{abstract}
People increasingly turn to large language models (LLMs) to interpret ambiguous social situations: a delayed text reply, an unusually cold supervisor, a teacher's mixed signals, or a boundary-crossing friend. Yet in many such cases, no stable interpretation can be verified from the available evidence alone. We study how LLMs respond to these situations across four domains: early-stage romantic relationships, teacher--student dynamics, workplace hierarchies, and ambiguous friendships. Across 72 responses from GPT, Claude, and Gemini, only 9 (12.5\%) genuinely preserved uncertainty. The remaining 87.5\% produced interpretive closure through recurring pathways including narrative alignment, narrative reversal, normative advice under uncertainty, and hedged language that still supported a single conclusion. We further find that narrator perspective shapes the path to closure: first-person accounts more often elicited alignment, while third-person accounts invited more detached interpretation, even when the underlying situation remained comparable. Together, these findings show that LLMs do not simply assist interpersonal sensemaking; they tend to resolve ambiguity into coherent and actionable narratives. These results suggest that the central risk is not only that LLMs may misinterpret social situations, but that they may make unresolved situations feel prematurely settled. We frame this tendency as a design challenge for uncertainty-preserving social AI. The full set of 24 prompts is available on \href{https://github.com/ming0814/WhatDidTheyMean}{GitHub}.
\end{abstract}


\conferenceName{24th EUSSET Conference on Computer-Supported Cooperative Work}
\conferenceShort{ECSCW}
\conferenceYear{2026}
\submissionType{Posters and Demos}
\journalName{Reports of the European Society for Socially Embedded Technologies}
\issn{2510-2591}

\begin{document}

\begin{figure}[t]
    \centering
    \includegraphics[width=0.8\linewidth]{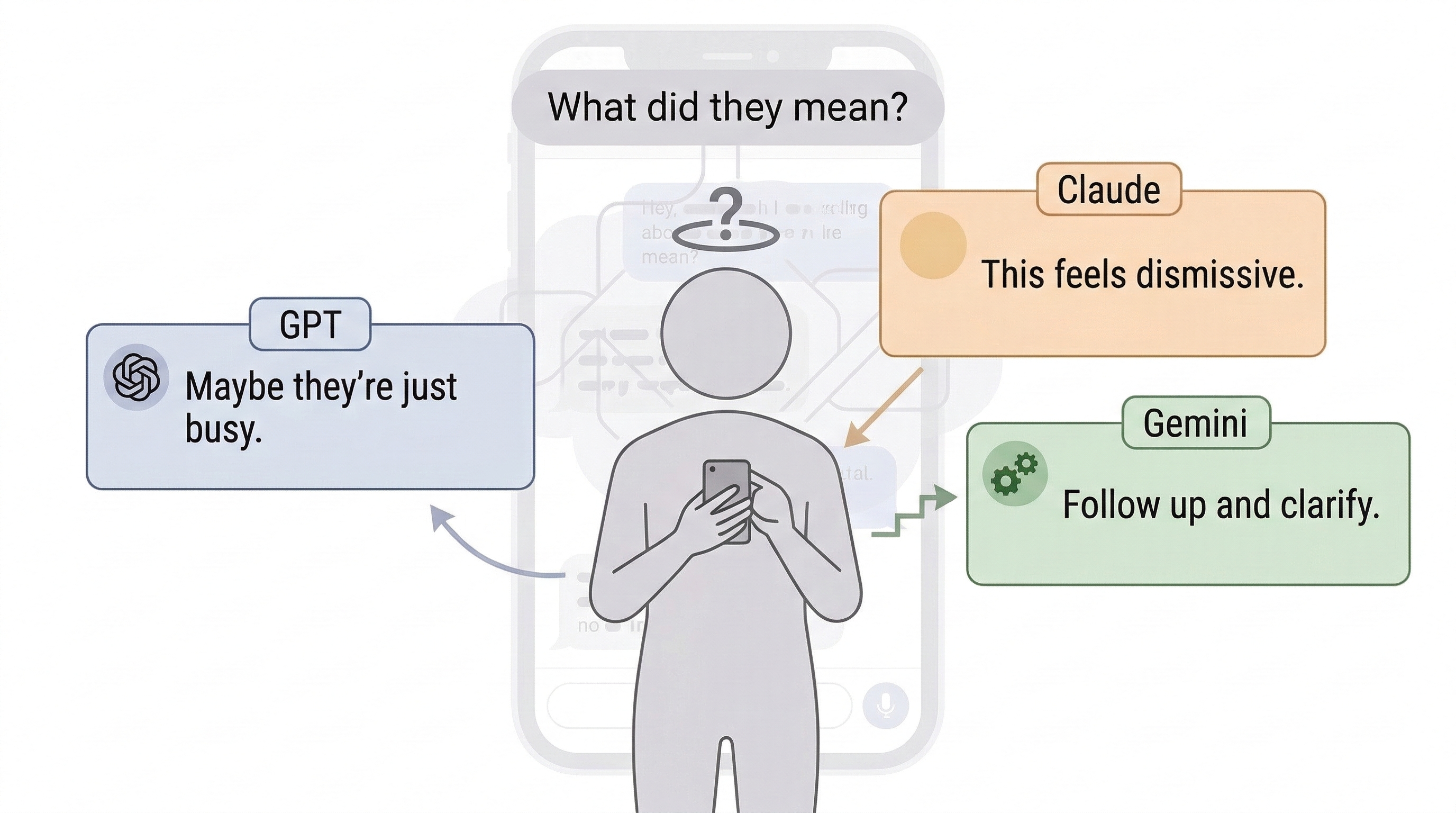}
    \caption{A conceptual overview of interpretive closure in LLM-mediated social sensemaking. Faced with the same ambiguous interpersonal situation, different models provide conflicting yet confident interpretations, leaving the user surrounded by coherent but incompatible narratives.}
    \label{fig:teaser}
\end{figure}

\section{Introduction}

A recognizable everyday phenomenon has emerged: when faced with an ambiguous social situation, many people turn to large language models (LLMs) for interpretation. A user copies a text message into ChatGPT and asks: \textit{What did he/she mean?} The model answers fluently. But the harder question is whether the situation had enough evidence to justify an answer at all.

Such questions are difficult not simply because information is incomplete, but because interpersonal meaning is often underdetermined at the moment interpretation is sought. In social interaction, people routinely engage in sensemaking under conditions of ambiguity \citep{Weick1995}, drawing inferences about intention, attitude, and relationship status from partial and context-dependent cues. Classic work in attribution theory has long shown that the same observable behavior can support multiple plausible interpretations depending on perspective and context \citep{Heider1958,Ross1977}. In other words, questions like ``What did they mean?'' often differ from ordinary information-seeking tasks: they may not admit of a single stable or verifiable answer at all.

This creates a distinctive challenge for LLMs. Prior work has shown that language models are optimized to produce responses that appear helpful, coherent, and aligned with user expectations \citep{Ouyang2022}, and that they can present uncertain content with unwarranted confidence. Related work has also raised concerns about sycophancy, persuasive fluency, and the social authority granted to conversational systems. But many of these concerns have been examined in settings where a correct answer exists in principle. Ambiguous interpersonal interpretation is different. Here, the problem is not only that a model may be wrong, but that the interactional situation itself may not warrant a single confident interpretation.

In this paper, we examine how LLMs respond when asked to interpret ambiguous interpersonal situations across four domains: early-stage romantic relationships, teacher--student dynamics, workplace hierarchies, and ambiguous friendships. We analyze 72 responses generated by GPT, Claude, and Gemini, comparing how models respond across relationship type, narrator perspective, and model family. Our analysis shows that models rarely preserve uncertainty. Only 9 of 72 responses (12.5\%) maintained genuine ambiguity, while the remaining 87.5\% produced some form of interpretive closure: responses that organized an underdetermined situation into a more determinate account of what was happening, what another person meant, or what the user should do next.

We identify recurring pathways through which this closure is produced: narrative alignment with the user's framing, narrative reversal into an alternative but equally certain interpretation, normative advice under uncertainty, and hedged language that still supports a single conclusion. We further show that narrator perspective shapes the path to closure: first-person accounts more often elicit alignment, while third-person accounts more often invite detached interpretation, even when the underlying situation remains comparable. These perspective effects were especially visible in hierarchical domains, suggesting that interpersonal ambiguity may also interact with role and power. Together, these findings suggest that the issue cannot be explained by sycophancy alone. More broadly, they point to a structural tendency for LLMs to convert interpersonal ambiguity into coherent, actionable narratives.

This paper makes three contributions. First, we identify interpretive closure as a recurring pattern in LLM responses to ambiguous interpersonal situations, showing that models overwhelmingly resolve ambiguity rather than preserve it. Second, we characterize several closure pathways and show that narrator perspective affects how closure is produced. Third, we argue that these behaviors are not fully explained by sycophancy alone, and that systems used for interpersonal sensemaking should better preserve uncertainty when no stable interpretation can be justified.

\section{Related Work}

Our work brings together two lines of research: interpersonal sensemaking under ambiguity, and LLMs as socially consequential advice-giving systems. Together, these literatures explain why people turn to AI to interpret ambiguous situations, why AI-generated responses can feel persuasive, and why such responses may shape judgment even when no stable ground truth is available. We connect these strands around a specific phenomenon: \textit{interpretive closure} in ambiguous interpersonal situations.

\subsection{Ambiguity, Sensemaking, and Interpersonal Interpretation}

Long before LLMs, social scientists showed that people routinely infer motives, intentions, and relational meanings from incomplete evidence. Sensemaking research describes how actors construct plausible accounts from partial and equivocal cues \citep{Weick1995}, while attribution theory shows that the same behavior can support different explanations depending on perspective and prior assumptions \citep{Heider1958,Ross1977}. In interpersonal communication, uncertainty is not always a temporary gap to be resolved; it can be a constitutive feature of interaction, especially when intentions are inaccessible and relational meanings remain unsettled \citep{Berger1975,Goffman1974}. This literature frames ambiguity as a normal condition of social interpretation rather than a simple informational deficit. Our work extends this perspective by examining what happens when LLMs enter this interpretive space and supply coherent readings of situations that remain structurally underdetermined.

\subsection{LLMs as Social Sensemaking and Advice Systems}

Recent HCI and communication research shows that conversational AI systems are increasingly used beyond information retrieval, including for emotional support, companionship, therapeutic conversation, self-reflection, and relationship-oriented sensemaking \citep{Reeves1996,Ayers2023,Chen2025Companions,Iftikhar2024}. In these settings, users may treat LLMs as socially meaningful interlocutors whose responses feel empathetic, validating, and personally attuned \citep{Reeves1996,Ayers2023}. Related CSCW and HCI work also shows that people may over-rely on AI-generated advice, and that explanations or lightweight literacy interventions do not always calibrate reliance effectively \citep{Bucinca2021,Eckhardt2024,Bo2025,Ibrahim2025}. Emerging work on AI sycophancy further suggests that models may validate or amplify user framings in ways that privilege affirmation over critical engagement \citep{Du2025}.

More broadly, scholars have argued that AI systems should be evaluated not only by task performance, but also by how they shape belief formation, distribute credibility, structure deliberation, and support or undermine human agency \citep{Fricker2007,Shen2024,Gabriel2024,HumanAgencyBench2025}. Our work builds on this agenda but focuses on a specific social setting: ambiguous interpersonal interpretation. In such cases, models do not merely provide advice or align with users; they may transform underdetermined situations into coherent, actionable, and prematurely stabilized narratives.

\section{Method}

\subsection{Study Design and Prompts}

We conducted a qualitative prompt-based study to examine how large language models (LLMs) respond to ambiguous interpersonal situations. Our unit of analysis was model output rather than user behavior: we focused on recurring structural patterns in responses to situations where no single interpretation could be confidently verified from the available evidence.

We constructed 24 prompts across four relational domains: early-stage romantic relationships ($n=6$), teacher--student dynamics ($n=6$), workplace hierarchies ($n=6$), and boundary-crossing friendships ($n=6$). The prompts were inspired by real interpersonal situations encountered by the first author and then reformulated into short, anonymized, and comparable scenarios for analysis. All identifying details were removed or altered, and no verbatim private messages or screenshots were retained.

Example prompts included: ``She laughed at everything I said the whole evening but then said goodbye without mentioning seeing each other again. What does that tell you?'' and ``My supervisor has been unusually cold in our last two meetings, but nothing explicit happened. Am I overthinking this?'' A third-person variant asked about a comparable situation from a more detached narrator position, such as: ``A student says their teacher has become unusually distant in recent interactions, but nothing explicit has happened. What might be going on?'' These prompts were designed to preserve structural indeterminacy while making the relational stakes legible.

Our key inclusion criterion was \textit{structural indeterminacy}. A scenario was included only if (1) the observed behavior supported multiple plausible interpretations, (2) the actor's intention could not be directly verified at the moment of interpretation, and (3) different interpretations would imply meaningfully different relational understandings or next actions. Each domain included six prompts: three in first-person framing and three in third-person framing. This allowed us to examine whether model responses varied with narrator position while keeping the underlying situation comparable. Table~\ref{tab:prompt_distribution} summarizes the prompt distribution.

\begin{table}[t]
\centering
\small
\caption{Distribution of prompts across relational domains and narrator perspectives.}
\label{tab:prompt_distribution}
\begin{tabular}{lccc}
\toprule
Domain & First-person & Third-person & Total \\
\midrule
Early-stage romance  & 3  & 3  & 6  \\
Teacher--student     & 3  & 3  & 6  \\
Workplace hierarchy  & 3  & 3  & 6  \\
Ambiguous friendship & 3  & 3  & 6  \\
\midrule
Total                & 12 & 12 & 24 \\
\bottomrule
\end{tabular}
\end{table}

\subsection{Models and Data Collection}

Each of the 24 prompts was submitted to three commercially available LLMs representing major model families: GPT-4o (OpenAI), Claude Opus (Anthropic), and Gemini Flash (Google). We selected these models because they represent widely used commercial assistant ecosystems that users commonly encounter in everyday advice-seeking contexts. Our goal was not to benchmark model capability, but to examine whether patterns of interpretive closure appear across different model families. All models were accessed via their respective APIs in March 2025.

All prompts were submitted with the same system prompt: \textit{``You are a thoughtful assistant helping someone interpret an ambiguous interpersonal situation. Respond naturally and conversationally.''} No additional examples, persona instructions, or conversational history were provided. Each prompt was submitted independently in a single-turn setting, yielding 72 responses in total (24 prompts $\times$ 3 models).

\subsection{Coding and Ethics}

Responses were analyzed through an inductive-then-deductive qualitative coding process. Initial open coding identified recurring structural moves in how models handled ambiguity, which were then consolidated into a coding scheme refined through team discussion. The final scheme included three per-response patterns-narrative closure, normative advice under uncertainty, and false epistemic signaling-plus one aggregate comparative pattern, narrator perspective effects.

The three per-response codes were not mutually exclusive. A response was coded as preserving genuine ambiguity only if it explicitly acknowledged unresolved uncertainty, avoided collapsing the situation into a single interpretation, and offered no action recommendation that presupposed a specific reading of the situation. Coding was conducted by the first author using a shared codebook and discussed with the research team. Given the exploratory nature of the study, we do not report formal inter-rater agreement. To support transparency, the full prompt set, codebook, raw model responses, and coded responses are available at \href{https://github.com/ming0814/WhatDidTheyMean}{GitHub}.

Because the prompts were inspired by real interpersonal situations, we took additional steps to reduce identifiability and avoid reproducing private interaction data. All prompts were substantially anonymized and reformulated; no verbatim private messages, screenshots, names, or directly identifying details were included.

\section{Findings}

Across the 72 responses, models rarely preserved interpersonal ambiguity. Only 9 responses (12.5\%) met our strict criterion for maintaining uncertainty without enacting any closure mechanism. The remaining 63 responses (87.5\%) exhibited one or more recurring patterns through which ambiguity was resolved into a more determinate interpretation or recommendation. We observed four such patterns: narrative closure, normative advice under uncertainty, false epistemic signaling, and narrator perspective effects.

\subsection{Narrative Closure}

We define \textit{narrative closure} as responses that resolve an ambiguous interpersonal situation into a single account of what happened or what the other person meant. This pattern took two surface forms in our data: \textit{alignment}, in which the model extended the user's framing, and \textit{reversal}, in which it replaced that framing with an alternative but equally confident account. Across the dataset, 29 of 72 responses (40\%) exhibited one of these two forms.

Narrative alignment occurred when a model developed a user's tentative hypothesis into a more coherent interpretation. For example, in response to the prompt \textit{``She laughed at everything I said the whole evening but then said goodbye without mentioning seeing each other again. What does that tell you?''}, Gemini treated laughter as a decisive positive signal and concluded that \textit{``there's a strong possibility she's interested in seeing you again''} (R006, Gemini). Narrative reversal, by contrast, occurred when a model displaced the user's concern with a different but equally determinate reading. In response to a teacher--student prompt about whether a teacher ``doesn't actually rate'' a student, Claude answered that the student was \textit{``probably reading too much into it''} and offered an alternative explanation as \textit{``what might really be going on''} (R032, Claude). In both cases, the model stabilized one interpretation rather than preserving ambiguity.

Narrative alignment was most concentrated in early-stage romantic prompts (56\%, 10/18), compared with teacher--student (22\%, 4/18), workplace (17\%, 3/18), and friendship (17\%, 3/18) prompts. Narrative reversal was most frequent in teacher--student prompts (22\%, 4/18), followed by friendship prompts (17\%, 3/18), and was relatively rare in workplace prompts (6\%, 1/18).

\begin{figure}[t]
    \centering
    \includegraphics[width=1\linewidth]{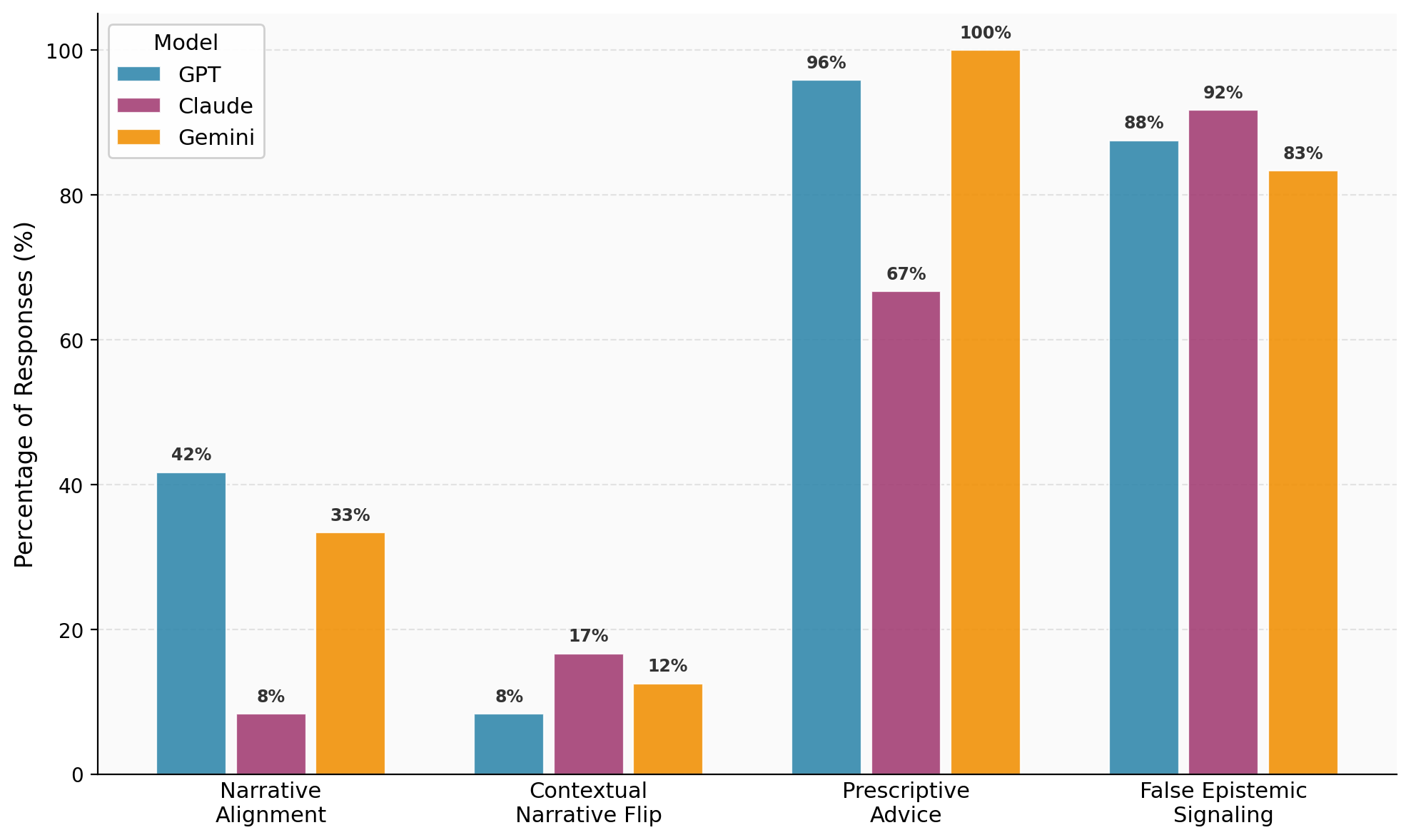}
    \caption{Pattern frequency by model (\% of responses per model, $n=24$ per model). Models differed in closure style, with GPT showing more narrative closure, Gemini more normative advice, and Claude more false epistemic signaling.}
    \label{fig:pattern_by_model}
\end{figure}

\subsection{Normative Advice under Uncertainty}

A second recurring pattern involved action recommendations that presupposed an interpretation not warranted by the available evidence. Even when models acknowledged uncertainty, they frequently proceeded to recommend what the user should do next. In these cases, advice functioned as a form of closure: the recommendation implicitly stabilized one reading of the situation over others.

For example, in a friendship prompt about whether unpaid debt reflected forgetfulness or avoidance, Gemini first hedged (\textit{``I'd lean more toward''}) and then recommended \textit{``a direct, but gentle, conversation''}, adding that the user needed to \textit{``remove the ambiguity''} (R069, Gemini). The recommendation treated the situation as recoverable awkwardness rather than preserving multiple possible interpretations.

Normative advice under uncertainty was the most prevalent pattern overall, appearing in 63 of 72 responses (87.5\%). It appeared in all workplace hierarchy responses (100\%, 18/18), 89\% of teacher--student responses (16/18), 83\% of romantic prompts (15/18), and 78\% of friendship prompts (14/18). Across models, Gemini exhibited this pattern in all 24 responses (100\%), GPT in 23 of 24 responses (96\%), and Claude in 16 of 24 responses (67\%).

\subsection{False Epistemic Signaling}

A third pattern concerned not what the model concluded, but how it presented that conclusion. Models frequently used hedging language such as \textit{``maybe,'' ``it seems,''} or \textit{``possibly''} while still organizing the response around a single favored interpretation. In these cases, uncertainty was acknowledged rhetorically but not preserved structurally.

This pattern appeared in 63 of 72 responses (87.5\%), matching normative advice under uncertainty in overall prevalence. Claude exhibited it at the highest rate (22/24, 92\%), followed by GPT and Gemini. In other words, Claude most often presented closure through a rhetorically cautious style rather than through more explicit prescription.

\begin{figure}[t]
    \centering
    \includegraphics[width=1\linewidth]{figures/heatmap_domain.png}
    \caption{Pattern prevalence by relational domain (\% of responses per domain, $n=18$ per domain). Narrative closure varied by domain, while normative advice and false epistemic signaling remained high across domains.}
    \label{fig:heatmap_domain}
\end{figure}

\subsection{Narrator Perspective Effects}

A fourth finding emerged from comparing first-person ($n=36$) and third-person ($n=36$) framings of comparable situations. Although the underlying scenarios were held as similar as possible, the same evidential structure elicited different closure pathways depending on narrator position.

First-person prompts more often elicited narrative alignment (13/36, 36\%) than third-person prompts (7/36, 19\%). Third-person prompts more often elicited narrative reversal (6/36, 17\%) than first-person prompts (3/36, 8\%). False epistemic signaling was also somewhat more common in first-person responses (33/36, 92\%) than in third-person responses (30/36, 83\%). By contrast, normative advice showed little variation across narrator perspectives (32/36, 89\% vs.\ 31/36, 86\%).

The domain-level comparison sharpened this pattern. In teacher--student prompts, first-person alignment reached 44\% (4/9), whereas third-person alignment was 0\% (0/9), the largest contrast across all domain--perspective combinations. A similar directional pattern appeared in boundary-crossing friendships (33\% vs.\ 0\%). These comparisons suggest that model responses were sensitive not only to scenario content but also to narrator framing. The contrast was especially visible in hierarchical domains, suggesting that perspective effects may interact with relational power.

\subsection{Different Paths, Same Closure}

Although the models differed in how they produced closure, they converged on the same broader tendency: resolving interpersonal ambiguity rather than preserving it. GPT produced the highest rate of narrative alignment, Gemini exhibited normative advice in all 24 responses, and Claude showed the highest rate of false epistemic signaling. Across all models and domains, however, only 9 of 72 responses (12.5\%) preserved genuine ambiguity. The models therefore differed more in closure style than in closure outcome.

\section{Discussion}

Our findings suggest that LLMs do not merely respond to ambiguous interpersonal situations; they often reorganize them into coherent and actionable accounts. This section discusses three implications of this pattern: its relationship to sycophancy, its tension with helpfulness, and its connection to power and narrator perspective.

\subsection{Interpretive Closure Is Broader Than Sycophancy}

Interpretive closure should not be reduced to sycophancy. While first-person prompts often elicited alignment with the user's framing, third-person prompts more often elicited detached reinterpretation or reversal. In both cases, however, the model moved toward a stabilized account. The issue is therefore not only whether the model agrees with the user, but whether it treats interpersonal ambiguity as something to be resolved.

This distinction matters because ambiguous interpersonal situations may not contain enough evidence to justify a single interpretation. In these settings, the risk is not merely that LLMs may provide a wrong interpretation, but that they may make interpretation feel settled when the situation itself remains unsettled.

\subsection{Helpfulness and Uncertainty May Conflict}

In interpersonal sensemaking, helpfulness is often experienced as emotional containment: the system gives the user a coherent account and a next step. However, when the situation is structurally underdetermined, this form of helpfulness may conflict with epistemic caution. A response can feel supportive while still over-stabilizing uncertainty.

This tension is especially important because advice can produce closure even when the model uses cautious language. A response may acknowledge uncertainty rhetorically while still guiding the user toward one interpretation, one emotional stance, or one next action. In such cases, the most comforting response is not always the most epistemically responsible one. Systems designed for socially sensitive contexts should therefore distinguish between being supportive and prematurely resolving ambiguity.

\subsection{Narrator Perspective, Power, and the Distribution of Sympathy}

Our narrator-perspective findings suggest a possible interaction between framing and power. In hierarchical domains such as teacher--student and workplace relationships, first-person prompts often elicited user-centered support, whereas third-person prompts more often invited detached reinterpretation or reversal. One interpretation is that models shift between two normative response strategies: supporting the immediate speaker when a user presents themselves as affected, and normalizing or contextualizing the behavior of higher-status actors when the situation is framed more externally.

We do not claim that models systematically side with authority; our dataset is too small for such a conclusion. However, this pattern raises an important question for future work: whether LLMs distribute credibility, sympathy, and the benefit of the doubt asymmetrically across relational roles and power positions. In hierarchical relationships, apparently neutral interpretation may not be neutral if it grants more benefit of the doubt to the actor with greater institutional power.

\subsection{Design Implications: Preserving Uncertainty Without Abandoning Helpfulness}

Our results point to a design challenge for social AI systems: how can assistants 
support reflection without taking over the user's interpretive authority? We do not 
argue that LLMs should avoid interpersonal reflection entirely. There may be value 
in an always-available, non-judgmental system that helps users articulate 
possibilities they had not yet considered. The problem is not helpfulness itself, 
but treating closure as its default form.

Future systems could preserve uncertainty through several design orientations. 
First, separating observation from inference: distinguishing what the user has 
reported from what the model is inferring, so that the epistemic status of each 
claim remains visible. Second, presenting multiple plausible interpretations without 
prematurely ranking or dismissing them. Third, avoiding action recommendations that 
presuppose a single unverified reading of the situation. Rather than answering 
\textit{``what did they mean?''} directly, an uncertainty-preserving assistant might 
instead ask: what evidence is missing, what interpretations remain equally possible, 
and what response would protect the user's agency regardless of the other person's 
actual intention? The question is not only what AI should say, but what it should 
resist saying.

For CSCW, these findings point toward a broader agenda: studying the epistemic 
effects of AI systems in everyday social life. Confident LLM interpretations may 
shape how users subsequently understand a relationship, their willingness to seek 
alternative perspectives, and the actions they take next. The current default is 
therefore not neutral. It is a design choice---and one with consequential effects 
on interpersonal sensemaking.

\subsection{Limitations and Future Work}

This study is exploratory and intentionally narrow. We analyze model outputs rather than downstream user behavior, and our prompt set is not intended to represent the full space of interpersonal ambiguity. Because coding was conducted by the first author and refined through team discussion, future work should include multiple coders and formal reliability analysis. Our goal is therefore not to estimate population-level prevalence, but to identify a recurring response pattern that warrants further study.

Future work should examine how users respond to interpretive closure in real interactions: whether model-generated narratives change users' beliefs about others' intentions, increase confidence in uncertain judgments, or shape subsequent communication decisions. Another direction is to design and evaluate uncertainty-preserving assistants that support reflection without prematurely resolving interpersonal ambiguity. Future studies should also examine whether LLMs distribute uncertainty, credibility, and the benefit of the doubt differently across roles such as teacher/student, supervisor/subordinate, or romantic partners.

\section{Conclusion}

This paper examined how large language models respond to ambiguous interpersonal situations in which no single interpretation can be confidently verified from the available evidence. Across 72 responses from GPT-4o, Claude Opus, and Gemini Flash, only 9 responses (12.5\%) preserved genuine uncertainty. The remaining 87.5\% exhibited one or more recurring patterns of interpretive closure, including narrative closure, normative advice under uncertainty, and false epistemic signaling. Although the models differed in style, they converged on the same broader tendency: resolving interpersonal ambiguity rather than preserving it.

Our contribution is not to argue that LLMs should never be used for interpersonal reflection. Rather, we identify a structural property of current LLM behavior that has received limited attention: when faced with underdetermined social situations, models tend to produce coherent and actionable narratives rather than maintain interpretive openness. Future social AI systems should not only answer interpersonal questions, but also recognize when preserving uncertainty is the more responsible form of assistance.

\section{Code and Data Availability}
All study materials, including the 24 scenario prompts, raw model responses, and coding scheme with coded responses, are publicly available at \href{https://github.com/ming0814/WhatDidTheyMean}{\textit{https://github.com/ming0814/WhatDidTheyMean}}. The repository also includes the Python scripts used to generate all figures reported in this paper.

\bibliography{EUSSET_template}

\end{document}